# О НЕКОТОРЫХ ОПТИМИЗАЦИОННЫХ ЗАДАЧАХ НА ГРАФАХ, СВОБОДНЫХ ОТ ЗВЁЗД

В.Г. Найденко, Ю.Л. Орлович

Беларусь, г. Минск

В основном используется терминология и обозначения книг [1, 2]. Термин "*граф*" означает конечный неориентированный граф без петель и кратных рёбер. Обозначим через $\Gamma$ множество всех графов, различаемых с точностью до изоморфизма. Пусть $Z \subseteq \Gamma$. Определим $FIS(Z)$ как множество всех таких графов, которые не имеют порождённых подграфов, принадлежащих $Z$. Граф $G$ назовём $K_{1,r}$-*свободным*, если $G \in FIS(K_{1,r})$. Через $n$, $\alpha(G)$, $\gamma(G)$, $\gamma_0(G)$ обозначаются соответственно порядок, число независимости, число доминирования и число вершин в наименьшем доминирующем независимом множестве графа $G$.

В дальнейшем, под *NP-оптимизационной проблемой* будем понимать четвёрку $Q = (I_Q, F_Q, f_Q, opt)$, где:

1) $I_Q$ – множество индивидуальных примеров для $Q$;

2) $F_Q(I)$ – множество допустимых решений для индивидуального примера $I \in I_Q$. Термин "*допустимый*" означает, что размер элементов $S \in F_Q(I)$ полиномиально ограничен размером примера $I$ и множество $\{(I,S) : S \in F_Q(I)\}$ распознаваемо за полиномиальное время;

3) $f_Q : \{(I,S) : S \in F_Q(I)\} \to N$ – полиномиально-вычислимая функция, называемая также целевой функцией;

4) $opt \in \{\max, \min\}$.

Положим $opt_Q(I) = opt\, f_Q(I,S)$ по всем $S \in F_Q(I)$. *Погрешностью до-*

*пустимого решения* S *для индивидуального примера* $I \in I_Q$ называется величина

$$R(I,S) = \begin{cases} \frac{opt_Q(I)}{f_Q(I,S)}, \text{ если } opt = \max; \\ \frac{f_Q(I,S)}{opt_Q(I)}, \text{ если } opt = \min. \end{cases}$$

Пусть $\varepsilon: N \to (1; +\infty)$ – произвольная функция. Будем говорить, что *NP-оптимизационная проблема Q аппроксимируется с точностью до фактора $\varepsilon$*, если существует полиномиальный алгоритм П который для каждого индивидуального примера $I \in I_Q$ отыскивает допустимое решение $S \in F_Q(I)$ с погрешностью $R(I,S) \le \varepsilon(|I|)$. Алгоритм П в этом случае называется *$\varepsilon$-приближённым алгоритмом* решения задачи Q. Класс всех NP-оптимизационных проблем, аппроксимируемых с точностью до постоянного фактора, обозначим через *APX*.

Отметим, что исследование конкретных NP-трудных оптимизационных проблем естественно проводить наряду с исследованием принадлежности их к классу *APX*, так как именно последнее обстоятельство является важным с точки зрения разработки эффективных приближённых алгоритмов для этих задач на практике.

В настоящей работе рассматриваются следующие три NP-оптимизационные проблемы на произвольном графе $G \in FIS(K_{1,r})$, $r \ge 3$.

1. *Наименьшее доминирующее множество (НДМ)*. Найти такое подмножество $V \subseteq VG$ наименьшей мощности, что каждая вершина из $VG \setminus V$ смежна с некоторой вершиной из V.

2. *Наибольшее независимое множество (ННМ)*. Найти такое подмножество $V \subseteq VG$ наибольшей мощности, что никакие две вершины из V не смежны.

3. *Наименьшее доминирующее независимое множество (НДНМ)*. Най-

ти подмножество $V \subseteq VG$ наименьшей мощности, которое является как независимым, так и доминирующим.

Отметим, что в классе всех графов указанные выше проблемы весьма трудны для аппроксимации. Действительно, пусть $v$ – произвольное число из интервала $(0;1)$. Тогда, как известно из [3], НДМ не аппроксимируется с точностью до $(1-v)\log n$, если $NP \not\subset DTIME(n^{\log\log n})$. Утверждение о невозможности аппроксимации ННМ с точностью до $n^{1-v}$ было получено в [4] при условии, что $coRP \neq NP$. Невозможность аппроксимации задачи НДНМ с точностью до $n^{1-v}$ доказана в [5].

В такой ситуации определённый интерес представляет рассмотрение задач НДМ, ННМ и НДНМ в классе $FIS(Z)$ для некоторого $Z \subset \Gamma$. Особенно примечательны в этом плане те нетривиальные $Z$, для которых перечисленные выше проблемы аппроксимируются с точностью до постоянного фактора. В нашем случае выбор класса $FIS(K_{1,r})$ продиктован стремлением охватить по возможности большое семейство графов, а также приложениями, на которые ориентирована теория.

Для дальнейшего важна следующая

**Лемма 1.** *Для любого графа $G \in FIS(K_{1,r})$, $r \geq 2$ верны неравенства*

$$\gamma(G) \leq \alpha(G) \leq (r-1) \cdot \gamma(G).$$

Заметим, что верхняя оценка для числа независимости, устанавливаемая предыдущей леммой, достижима. Так, для графа $G = pK_{1,r-1}$ с параметрами $p \geq 1$, $r \geq 3$, имеем $\gamma(G) = p$ и $\alpha(G) = p(r-1)$.

Пусть, далее, $G \in FIS(K_{1,r})$, $r \geq 2$ и $V'_\alpha$ – одно из максимальных по включению независимых множеств графа $G$. Положим $\alpha' = |V'_\alpha|$. При этих обозначениях из леммы 1 вытекает

**Лемма 2.** $\quad \dfrac{\alpha'}{r-1} \leq \gamma(G) \leq \gamma_0(G) \leq \alpha', \quad \alpha' \leq \alpha(G) \leq (r-1) \cdot \alpha'.$

Применяя лемму 2 получаем следующее утверждение.

**Теорема 1.** *Любое максимальное по включению независимое множество $V'_\alpha$ графа $G \in FIS(K_{1,r})$, $r \geq 3$, является допустимым решением для задач НДМ, ННМ и НДНМ с постоянной погрешностью $r-1$.*

Как известно, максимальное независимое множество в произвольном графе можно найти за полиномиальное время градиентным алгоритмом, например, минимально-степенным алгоритмом **Greedy** [1]. Таким образом, в случае, если $G \in FIS(K_{1,r})$, теорема 1 устанавливает принадлежность задач *НДМ*, *ННМ* и *НДНМ* к классу *APX*. В частности, для *НДНМ* мы получили ранее не известную оценку погрешности, с которой эта задача аппроксимируется в классе $\Gamma_{\Delta \leq c}$ графов, максимальная степень $\Delta$ которых ограничена некоторой константой $c \geq 3$. Действительно, из теоремы 1 в силу включения $\Gamma_{\Delta \leq c} \subset FIS(K_{1,c+1})$ вытекает, что существует полиномиальный алгоритм, оценка погрешности которого для задачи *НДНМ* в классе $\Gamma_{\Delta \leq c}$ равна $c$. Отметим, что распознавательные аналоги всех трёх изучаемых задач остаются *NP*-полными в $\Gamma_{\Delta \leq c}$ для любого фиксированного $c \geq 3$ [2].

Следующая теорема показывает, что при решении задачи *ННМ* в классе $FIS(K_{1,r})$ алгоритмом **Greedy** границу $r-1$ улучшить нельзя.

**Теорема 2.** *Пусть $\alpha'(G)$ – мощность максимального по включению независимого множества графа $G$, найденного алгоритмом **Greedy**. Тогда*

$$\sup_{G \in FIS(K_{1,r})} \left\{ \frac{\alpha(G)}{\alpha'(G)} \right\} = r - 1,$$

*где $r \geq 3$.*

Отметим, что теорема 1 переносится на более общие классы графов, которые допускают, в частности, порождённые подграфы $K_{1,r}$. Рассмотрим один из таких классов. В целях сокращения записи введём обозначение $G_v = G(N(v))$, где $N(v)$ – окружение вершины $v$ в графе $G$. Зафиксируем целое число $r \geq 3$. Назовём вершины $u$ и $v$ $r$-эквивалентными, если

$$\min\{\alpha(G_u), \alpha(G_v)\} > r - 1.$$

Будем говорить, что граф $G$ *почти $K_{1,r}$-свободный*, если множество всех $r$-эквивалентных вершин независимо в $G$ и

$$\max_{v \in VG}\{\gamma(G_v)\} \leq r - 1.$$

Класс всех таких графов обозначим через $ACF(K_{1,r})$. Очевидно, что

$$FIS(K_{1,r}) \subset ACF(K_{1,r}).$$

**Теорема 3.** *Любое максимальное по включению независимое множество графа $G \in ACF(K_{1,r})$, $r \geq 3$ является допустимым решением для задач НДМ, ННМ и НДНМ с постоянной погрешностью $(r-1)^2$.*

Остается открытым вопрос: существуют ли полиномиальные алгоритмы с меньшей, чем $r - 1$ погрешностью для решения задач *НДМ*, *ННМ* и *НДНМ* в классе $FIS(K_{1,r})$ для любого фиксированного $r$? Возможным ответом на этот вопрос является локальный поиск.

В заключение рассмотрим понятия, относящиеся к локальному поиску. Пусть множество $S$ представляет собой допустимое решение для индивидуального примера $I \in I_Q$ некоторой *NP*-оптимизационной проблемы $Q$ (под проблемой $Q$ будем понимать, в зависимости от контекста, *ННМ*, *НДНМ* или *НДМ*). Тогда функция $f_Q$ при фиксированном $I$ есть $|S|$.

Множество $L$ называется *улучшением* для $S$, если симметрическая разность $S \oplus L$ также является допустимым решением для $I$, причём $|S| < |S \oplus L|$, если $opt = \max$ и $|S| > |S \oplus L|$, если $opt = \min$. Пусть $t$ – натуральное число. Улучшение $L$ называется *t-улучшением*, когда $|S \cap L| \leq t - 1$, если $opt = \max$ и $|S \cap L| \leq t$, если $opt = \min$. Допустимое решение $S$ будем называть *t-локальным*, если для него не существует *t*-улучшения. Приведём общую схему *t*-локального поиска.

**Вход:** $I, t$.

**Выход:** $t$-локальное решение $S$ для $I$.

**Схема:**

**Шаг 1.** Произвольно выбрать $S$ из множества $F_Q(I)$.

**Шаг 2.** Найти $t$-улучшение $L$ для $S$. Если $L$ не найдено, то перейти к шагу 4.

**Шаг 3.** $S := S \oplus L$ и перейти к шагу 2.

**Шаг 4.** Стоп.

Отметим, что максимальное по включению независимое множество $V'_\alpha$ из формулировки теоремы 1 есть в точности 1-локальное решение для проблемы *ННМ*. Поэтому было бы перспективным исследовать $t$-локальный поиск для *НДМ*, *ННМ*, *НДНМ* в классе $FIS(K_{1,r})$ при $t > 1$. При этом, с ростом $t$ трудоёмкость поиска естественно возрастёт, однако появится возможность найти алгоритмы с более высокой точностью.

## Литература